\documentstyle[prd,aps]{revtex}

\begin{document}
\draft

\title{The Schrodinger Equation From a Quadratic Hamiltonian System}
\author{Wai Bong Yeung\footnote{Electronic address:
{phwyeung @ccvax.sinica.edu.tw}}}
\address{Institute of Physics, Academia Sinica, Taipei, Taiwan ,ROC}
\date{Oct 1999}
\maketitle

\begin{abstract}
We regard the real and imaginary parts of the Schrodinger wave
function as canonical conjugate variables.With this pair of
conjugate variables and some other 2n pairs, we construct a
quadratic Hamiltonian density. We then show that the Schrodinger
Equation follows from the Hamilton's Equation of motion when the
Planck frequency is much larger than the characteristic
frequencies of the Hamiltonian system. The Hamiltonian and the
normal mode solutions coincide, respectively, with the energy
expectation value and the energy eigenstates of the corresponding
quantum mechanical system.
\end{abstract}

Let us consider a Hamiltonian system consists of (1+2n) pairs of
canonical conjugate variables ($p,q$) ( $P_j , Q _j$), ($\pi_j
,\eta_j$  ), j=1,...,n.All of these canonical variables are
functionc of  $x=(x_1 ,.. ,x_n )$ and t.

The Hamiltonian density  H ($p,q;P_j,Q_j;\pi_j,\eta_j ;\partial
q,\partial Q,\partial\eta $) is quadratic in the variables and
chosen to be
\begin{equation}
 H=1/2( V(x)(p ^2 + q ^2 )-m(P_j^2   +Q_j^2  +\pi_j^2
 + \eta_j^2 )-p\partial  _j (Q _j+\eta_j )-(P _j +\pi_j    )\partial _j   q )
\end{equation}
where m is a given parameter of the system. The function V(x)is
also given. $\partial _j $  denotes the differentiation with
respect of$ x _j$ . Repeated indices  will be summed over from 1 to n.

The (2+4n) Hamilton Equation of motion ,obtained by varying
independently the canonical variables [1]  are
\begin{eqnarray} \partial _t p &= & -V(x)q -1/2 \partial _j  (
P_j+ \pi_j)           \nonumber\\
\nonumber\\
\partial_t q &= & V(x)p -1/2 \partial _j   (Q_j +\eta_j )
\end{eqnarray}
Similarly
\begin{eqnarray}
\partial _tP_i &=& mQ_i -1/2 \partial _i  p
\nonumber\\
\partial _tQ _i&= & -mp_ i-1/2 \partial _i   q
\nonumber\\                                       \partial
_t\pi_i& = & m\eta_i  -1/2 \partial _i  p
\nonumber\\
\partial _t\eta_i &= & -m \pi_i-1/2 \partial _i q
 \end{eqnarray}
These are coupled linear equations ,and hence linear
superpositions of solutions of these equations are also solutions to
these equations. In other words, the principle of linear superposition [2]
applies in this Hamiltonian system.   Now consider the case in which
m is a very large parameter and that  ($P_i $ ,$Q _i $ ) and
( $\pi_i $,$\eta_i $ ) are slowly varying variables . In that
case Eq(3) can be approximated by
\begin{eqnarray}
    Q_i &= & 1/2m \partial _i p        \nonumber\\
    P_i &= & -1/2m \partial _i q         \nonumber\\
          \eta_i & =& 1/2m \partial _i  p         \nonumber\\                                                         \pi_i&=& -1/2m \partial _i  q                               ,  j=1,..n                                                                                                    \end{eqnarray}
And because m is a very large number, ( $P_i ,Q_i $) and ($\pi_i  ,\eta_i $  )
 are small relative to ($p,q$). Eq(4) gives the relations between the large variables ($p,q$)
 and the small variables  ($P_i, Q_i $), ($\pi_i  ,\eta_i $ ).
                                                                                                                                     Eliminating the small variables from Eq(2)  , we get the following equations containiong the large variables only.
\begin{eqnarray}
 \partial _tp& =& -V(x) q +1/2m \partial^2  q\nonumber\\
 \partial _tq &= &V(x)p-1/2m \partial^2  p
  \end{eqnarray}
Let us now construct a complex fuction $\psi (x,t)$ by using the above $q(x,t)$
and $p(x,t) $as the real and imaginary parts of    $\psi(x,t)$=($q(x,t)+ip(x,t)$)/$\sqrt{2}$.
It can be seen immediately that Eq(5) can be summerized as the real and imaginary parts ,
respectively, of the complex equation
        \begin{equation}
i \partial_t\psi =(-1/2m \partial^2  +V(x))\psi
\end{equation}
 This complex equation is nothing but the Schrodinger equation in natural units (h=c=1).
 The parameter m and the function V(x ) will then be, respectively, the mass and
 the potential energy of the corresponding quantum mechanical system.

It is interesting to calculate the Hamiltonian $ H =\int  d ^n x H  $
for the system. With the relations given in Eq(4) substituded  into Eq(1),
the  Hamiltonian of the system will read as
     \begin{equation}
  H = \int d ^n x 1/2 (V(x)(p ^2+ q ^2 ) -m(1/2m ^2 (\partial_j  q )^2 +1/2m ^2 (\partial_j
    p)^2) -1/mp \partial^2  p  +1/m(\partial_j   q)^2 ) )                                                                                                    \end{equation}
Integration by parts and discarding surface terms, we get
\begin{equation}
 H=\int  d ^n x1/2(V(x)(p^2  +q ^2 )-p1/2m \partial^2  p -q1/2m \partial^2 q)                                               \end{equation}
Which is the same as $\int   d ^n x \psi* (-1/2m \partial^2   +V(x))\psi  $ .And
 hence the Hamiltonian  H of our Hamiltonian system is just the expectation value of
 the nonrelativistic energy of the corresponding wave function.

Furthermore ,Eq(5) can be used to rewrite the Hamiltonian H into the the form
\begin{equation}
  H= \int d ^nx 1/2(p\partial_tq -q\partial_tp)
\end{equation}
This H       could be a function of time even if $p(x,t) $and $q(x,t)$ satisfy Eq(5) If
 we are interested in looking for normal mode solutions for the Hamiltonian system ,
 then we should look for solutions which will give us an effective Hamiltonian of the form of
   \begin{equation}
  H =E\int  d ^n x1/2(p^2 +q^2 )
\end{equation}
Where E is a constant to be determined . Comparision of Eq(9)with Eq(10) will give
\begin{eqnarray}
   \partial_t q &=& Ep      \nonumber\\
   \partial _t p&=& -Eq                                                                                                    \end{eqnarray}
Eq(11) will ensure that H be time independent and will also cast Eq(5) into the form of
\begin{eqnarray}
   Eq &=& V(x) q  -1/2m \partial^2  q  \nonumber \\
   Ep &=& V(x) p-1/2m \partial^2 p
 \end{eqnarray}
This Eq(12) is exactly the same as the eigenvalue problem for
\begin{equation}
  E \psi =(V(x)-1/2m \partial^2   ) \psi
\end{equation}                                                                                                                                  To consider the Hamiltonian system in MKS units, we have to put in some dimensional parameters to take care of the different dimensionalities of various terms of the Hamiltonian density. If we choose p and q to be dimensionless , then the Hamilton's Equation of motion will require H to be 1/time in dimension . And since V(x) is the potential energy, the first term on the right hand side of Eq(10) must be divided by a constant of the dimension of energy-time in order to get the dimension right.This unique constant is, of course, the Planck's constant h .Simple arguments will fix the other constants before variuos terms , and the result is the following Hamiltonian density
\begin{equation}
  H=(1/2h)(V(x)(p^2 + q^2 )  -(1/2h)(mc ^2  )(P _j^2 + Q _j^2 + \pi_j^2  + \eta_j^2
   )-(c/2)p \partial_j  (Q _j + \eta_j )  -(c/2) ( P_j  +\pi_j  )\partial_j   q
\end {equation}
This Hamiltonian will generate the Schrodinger Equation
\begin{equation}
  ih\partial_t\psi   =( -h ^2/2m\partial^2    +V(x)) \psi
\end{equation}
under the conditions that we have mentioned above. And Eq(15), which in turn, will
imply the prescription
E $\longrightarrow ih\partial_t$  , momemtum $\longrightarrow -ih\partial_x$,
used in quantum mechanics for E$= (momentum)^2 /2m   + V(x)$.

The very large parameter will now be the Planck frequency $mc ^2
/h$. The low frequency sector of the Hamiltonian system , which is
the sector with characteristic
 frequecies considerably smaller than the Planck frequency , will coincide
  exactly with the Schrodinger quantum mechanics. Our restriction to the low
  frequency sector of the Hamiltonian system , we think , is not a drawback of
  our approach because of the fact that the Schrodinger quantum mechanics is a
nonrelaitivistic limit by itself and also that the Planck
frequency is always a gigiantic number for all known atomic
particle
In this approach of obtaining the Schrodinger Equation, two more
things worth our attention. Firstly, the Planck constant h appears
as a parameter in the Hamiltonian density. The role thus played by
h in our approach is radically different from that played in the
conventional presentation of quantum mechanics. Secondly
,physically there are the small hidden variable. They come from
the negative parts of the Hamiltonian density, and are playing
important roles in the determiantion of the evolution of $p(x,t)$
and $q(x,t)$, and hence $\psi  (x,t)$ [3] .

\end{document}